\documentclass[botnum, fleqn]{unmeethesis}
\usepackage{comment}
\usepackage{pdfpages}
\usepackage{xcolor}
\usepackage{algpseudocode}
\usepackage[font=footnotesize]{subfig}
\usepackage{amssymb}
\usepackage[cmex10]{amsmath}
\usepackage{graphicx}
\usepackage{rotating} % To rotate figure
\usepackage{caption} % to remove caption when we split figure onto multiple pages
\newcommand{\blu}{\color{blue}}         % Notes by Vj
\newcommand{\COMMENT}[1]{$\triangleright$  #1}
\newcommand{\C}[1]{$\triangleright$  #1}

\begin{document}
\frontmatter

% Uncomment the next command if you see weird paragraph spacing:
% That is, if you see paragraphs float with lots of white space
% in between them:

% \setlength{\parskip}{0.30cm}

\title{Automatic Segmentation of Coronal Holes in Solar Images and 
  Solar Prediction Map Classification}

\author{Venkatesh Jatla}

\degreesubject{M.Sc., Computer Engineering}

\degree{Master of Science \\ Computer Engineering}

\documenttype{Thesis}

\previousdegrees{B.Tech., Electronics and Communications, Vellore institute of Technology, 2007}
\date{December, 2016}
\maketitle
\setcounter{page}{3}
% \makecopyright
% Copyright page is no longer necessary D. Murrell

% DEDICAITON, no dedication for now
\begin{dedication}
	This thesis is dedicated to the Sun, by whose grace earth shines and whose
	warmth has always given me joy.   
\end{dedication}

\begin{acknowledgments}
  \vspace{1.1in}
  I would like to thank my advisor, Professor Marios S. Pattichis, for his support.
  I would like to thank Nick Arge, Samantha Demorco, Ratchel Hock and
  Callie Darsey for introducing the problem and manual annotations. I would
  also like to thank Andrew Delgado whose M.Sc. report provided a starting
  point to my research.
\end{acknowledgments}

\maketitleabstract %(required even though there's no abstract title anymore)

\begin{abstract}
  Solar image analysis relies on the detection of coronal holes
  for predicting disruptions to earth's magnetic field.
  The coronal holes act as sources of solar wind that can reach the earth.
  Thus, coronal holes are used in physical models for predicting
  the evolution of solar wind and its potential for 
  interfering with the earth's magnetic field.
  Due to inherent uncertainties in the physical models,
  there is a need for a classification system that
  can be used to select the physical models 
  that best match the observed coronal holes.
  
  The physical model classification problem is decomposed
  into three subproblems.
  First, the thesis develops a method for coronal hole segmentation.
  Second, the thesis develops methods for matching coronal holes from different maps.
  Third, based on the matching results,
  the thesis develops a physical map classification system.

  A level-set segmentation
  method is used  for detecting coronal holes
  that are observed in 
  extreme ultra-violet images (EUVI) and
  magnetic field images.
  For validating the segmentation approach,
  two independent manual segmentations
  were combined to produce 46 consensus maps.
  Overall, the level-set segmentation approach
  produces significant improvements over
  current approaches.

  Coronal hole matching is broken into two steps.
  First, an automated method is used
  to combine coronal holes into clusters.
  Second, a Linear Program formulation 
  is used for matching the clusters.
  The results are validated using manual clustering
  and matching.
  Compared to manual matching, the automated
  matching method gave more than 85 percent accuracy.
  
  Physical map classification is based on
  coronal hole matching between the physical maps and 
  (i)  the consensus maps (semi-automated), or
  (ii) the segmented maps (fully-automated).
  Based on the matching results, 
  the system uses area differences, shortest distances between matched clusters,
  number and areas of new and missing coronal hole clusters
  to classify each map.
  The results indicate that the automated segmentation and 
  classification system
  performs better than individual humans.    
  \clearpage %(required for 1-page abstract)
\end{abstract}

\tableofcontents
\listoffigures
\listoftables

\mainmatter
% 
% Chapter 1
% 
\chapter{Introduction}

\begin{figure}[bht]
  \centering
  \subfloat[EUVI image (synoptic image).]
  {\includegraphics[width=0.47\textwidth]{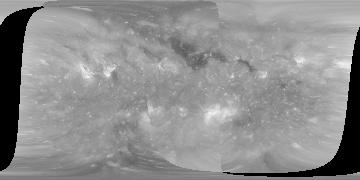}
    \label{subfig:detectionProblem_synopticImage}} 
  \subfloat[Magnetic image (photomap image).]
  {\includegraphics[width=0.47\textwidth]{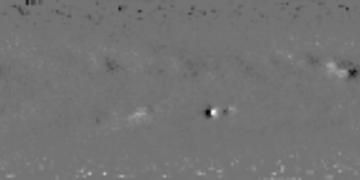}
    \label{subfig:detectionProblem_magneticImage}} \\
  \subfloat[Consensus map.]
  {\includegraphics[width=0.47\textwidth]{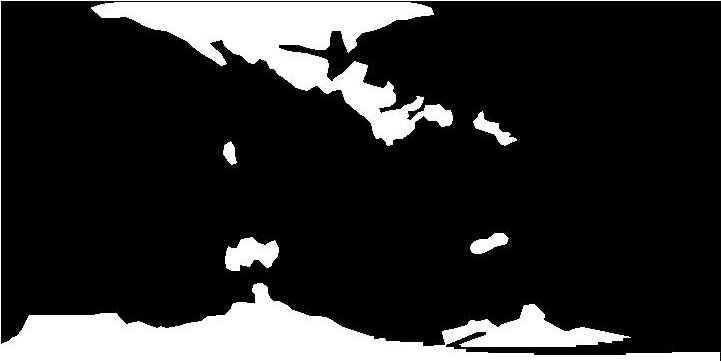}
    \label{subfig:detectionProblem_consensusImage}} 
  \subfloat[Current method: ${\tt unit\_dist} = 0.17$]
  {\includegraphics[width=0.47\textwidth]{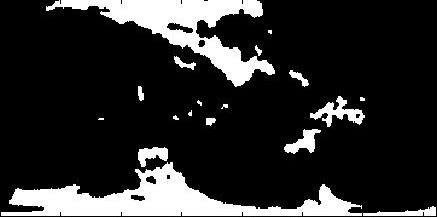}
    \label{subfig:detectinProblem_HenneyHarvey}}\\        
  \caption{\label{fig:detectionProblem}
    The basic coronal hole detection problem.
    Example input data maps for May 2, 2011 are shown:
    EUVI (a) and magnetic image in (b).
    The consensus map for the coronal holes is shown in (c).
    Results from using the current method described in \cite{Henney2005}
    can be found in (d).
    The {\tt unit\_dist} refers to the distance of the current algorithm from
    the ideal performance given by $({\tt sensitivity}, \, {\tt specificity})=(1, 1)$,
    where the ideal image is given in (c).    
  }         
\end{figure}

\begin{figure}
  \centering
  \subfloat[Consensus map]{
    \includegraphics[width=0.66\textwidth]{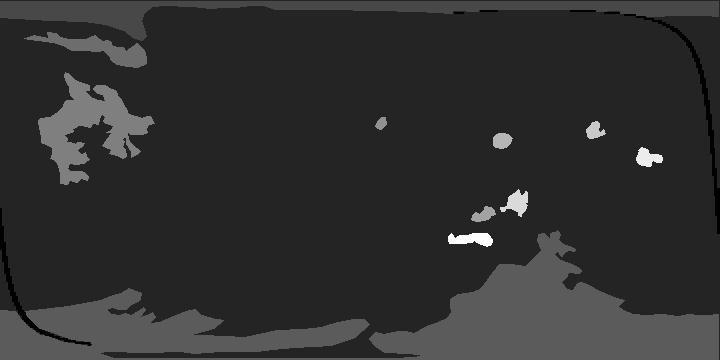}
    \label{subfig:classificaitonProblem_consensus}}
  
  \subfloat[Model 1]{
    \includegraphics[width=0.47\textwidth]{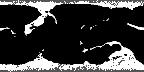}
    \label{subfig:classificaitonProblem_m1}}
  \subfloat[Model 2]
  {\includegraphics[width=0.47\textwidth]{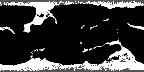}
    \label{subfig:classificaitonProblem_m2}}
  
  \subfloat[Model 3]
  {\includegraphics[width=0.47\textwidth]{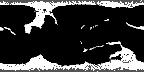}
    \label{subfig:classificaitonProblem_m3}}        
  \subfloat[Model 4]
  {\includegraphics[width=0.47\textwidth]{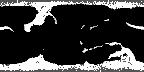}
    \label{subfig:classificaitonProblem_m4}}
  
  \subfloat[Model 5]
  {\includegraphics[width=0.47\textwidth]{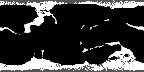}
    \label{subfig:classificaitonProblem_m5}}
  \subfloat[Model 6]
  {\includegraphics[width=0.47\textwidth]{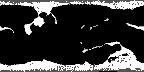}
    \label{subfig:classificaitonProblem_m6}}
  \caption{Physical model classification problem (I of II). 
    Physical models are classified based on their distance from the 
    consensus map. The maps correspond to 21-01-2011.}
  \label{fig:classificaitonProblem1} 
\end{figure}
\begin{figure}
  \centering              
  \subfloat[Consensus map]{
    \includegraphics[width=0.66\textwidth]{pictures/chapter1/classification/consensus.jpg}
    \label{subfig:classificaitonProblem_consensus}}
  
  \subfloat[Model 7]
  {\includegraphics[width=0.47\textwidth]{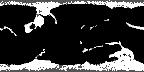}
    \label{subfig:classificaitonProblem_m7}}
  \subfloat[Model 8]
  {\includegraphics[width=0.47\textwidth]{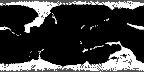}
    \label{subfig:classificaitonProblem_m8}}

  \subfloat[Model 9]
  {\includegraphics[width=0.47\textwidth]{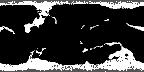}
    \label{subfig:classificaitonProblem_m9}}
  \subfloat[Model 10]
  {\includegraphics[width=0.47\textwidth]{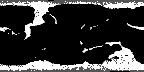}
    \label{subfig:classificaitonProblem_m10}}
  
  \subfloat[Model 11]
  {\includegraphics[width=0.47\textwidth]{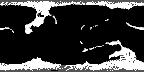}
    \label{subfig:classificaitonProblem_m11}}
  \subfloat[Model 12]
  {\includegraphics[width=0.47\textwidth]{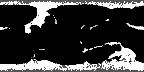}
    \label{subfig:classificaitonProblem_m12}}
  \caption{Physical model classification problem (II of II). 
    Physical models are classified based on their distance from the 
    consensus map. The maps correspond to 21-01-2011.}
  \label{fig:classificaitonProblem2} 
\end{figure}

Coronal holes represent regions of the  solar surface where the plasma density is low. 
Inside the coronal holes, the magnetic field 
lines are open and stretch into space. 
Thus, coronal holes can be sources of  solar winds that 
can escape at high speeds. 
Depending on its location, a solar wind that originates from a coronal hole
can reach the earth and cause large disturbances to the earth's magnetic field. If the disturbances are sufficiently large, they can cause significant damage to the power 
and communications infrastructure. To avoid such damage, it is important to develop reliable methods for predicting the locations and the sizes of the coronal holes.

Coronal holes can be observed in solar images taken in various spectral bands.
Based on the detected coronal holes, physical models can then be used to predict solar wind propagation. Unfortunately, there are strong variations in the physical models that can be used for forecasting. There is thus a strong need to reliably detect the coronal holes and select physical models that match the detected coronal holes to support accurate solar predictions.  A fundamental problem
associated with developing analysis methods comes from the lack of reliable maps that outline the locations of the coronal holes.\\

To assess the performance of methods for detecting coronal holes, there is a need
to establish a ground truth from solar observations. 
Unfortunately, the standard practice of comparing automatic and manual
annotations from a single expert is not acceptable since such an approach can lead to significant biases. A manual annotation tool developed by Dr.\ Pattichis in the Summer of 2012 has been used to allow multiple users to manually outline the locations and regions of coronal holes within specific dates. The system was used to generate coronal hole maps by two different users, and then again to derive  consensus annotations by combining the two independent maps. The consensus maps have the unique advantage that they avoid biases associated
with individual raters.

This thesis develops automated image processing models that can be used to detect coronal holes automatically (see Fig.  \ref{fig:detectionProblem}) and then classify physical models on whether they correctly match the coronal holes that have been detected. The basic approach relies on the detection of coronal holes based on  extreme ultra-violet images (EUVI) and magnetic images.
An automated segmentation method has been developed that improves significantly over a previously considered method described in \cite{Henney2005}.
The thesis also extends prior, initial work by Andrew Delgado  to address the matching problem.

To develop the necessary methods, the thesis developed and implemented manual classification
protocols that were used to establish a ground truth database for
the classification of physical models.
The manual classification is used to identify the best
physical models based on how well they match the consensus maps.

We present an example of the physical model classification problem in 
Figs. 
\ref{fig:classificaitonProblem1} and \ref{fig:classificaitonProblem2}.
The basic idea is to develop a classification system
that grades physical models based on their agreement
with the consensus map.
The physical maps that are closer to the consensus map are
classified as good for use in forecasting.
On the other hand, the maps that are significantly different
from the consensus map, are to be classified as
unsuitable for use for forecasting.

\section{\label{sec:thStatement}Thesis Statement}
My thesis is that I can develop an automated system
that can be used to classify physical models based on how well
they can be used to predict the location and areas of coronal holes 
that are themselves detected from solar images.

\section{\label{sec:contributions}Contributions}
The primary contribution of the thesis include:
\begin{itemize}
\item \textbf{\textit{Manual classification of physical models:}}
  The thesis provides a protocol for the manual classification
  of physical models based on how well they match coronal holes
  seen in consensus maps.
\item \textbf{\textit{Level sets segmentation of coronal holes:}}
  The thesis describes a hybrid method that uses magnetic images
  and EUVI to detect coronal holes.
\item \textbf{\textit{Coronal holes matching algorithm:}}
  The thesis describes a matching method that is used for matching 
  clusters of coronal holes between maps and also to detect
  new (generated) and missing (removed) coronal holes.    
\item \textbf{\textit{Computer classification of physical maps:}}
  An automated algorithm has been developed for selecting physical maps
  that should be used for solar prediction. The computer classification
  methods has been validated using manual classification.
\end{itemize}

\section{\label{sec:thOverview}Thesis Overview}
The remainder of the thesis is organized into 5 chapters:
\begin{itemize}
\item \textbf{Chapter 2: Background.} 
  This chapter describes prior work.
\item \textbf{Chapter 3: Manual classification of physical models.}
  This chapters describes the visual interface and the protocol that 
  was used to create a ground truth of manual classifications for the physical models.
\item \textbf{Chapter 4: Computer segmentation and classification of physical models.}
  This chapter describes the coronal hole segmentation, matching algorithm and the automated
  classification system that is used to select the best physical models.
\item \textbf{Chapter 5: Results.} This chapter provides a summary of the results
  for coronal hole segmentation, matching and computer classification, as compared to manual 
  classification.
\item \textbf{Chapter 6: Conclusion and future work.}  
  This chapter provides a summary of the thesis and recommendations
  for future work.                         
\end{itemize}

% Chapter background
\chapter{Background}
\section{Creating consensus maps}
\begin{figure}
  \includegraphics[width=\textwidth]{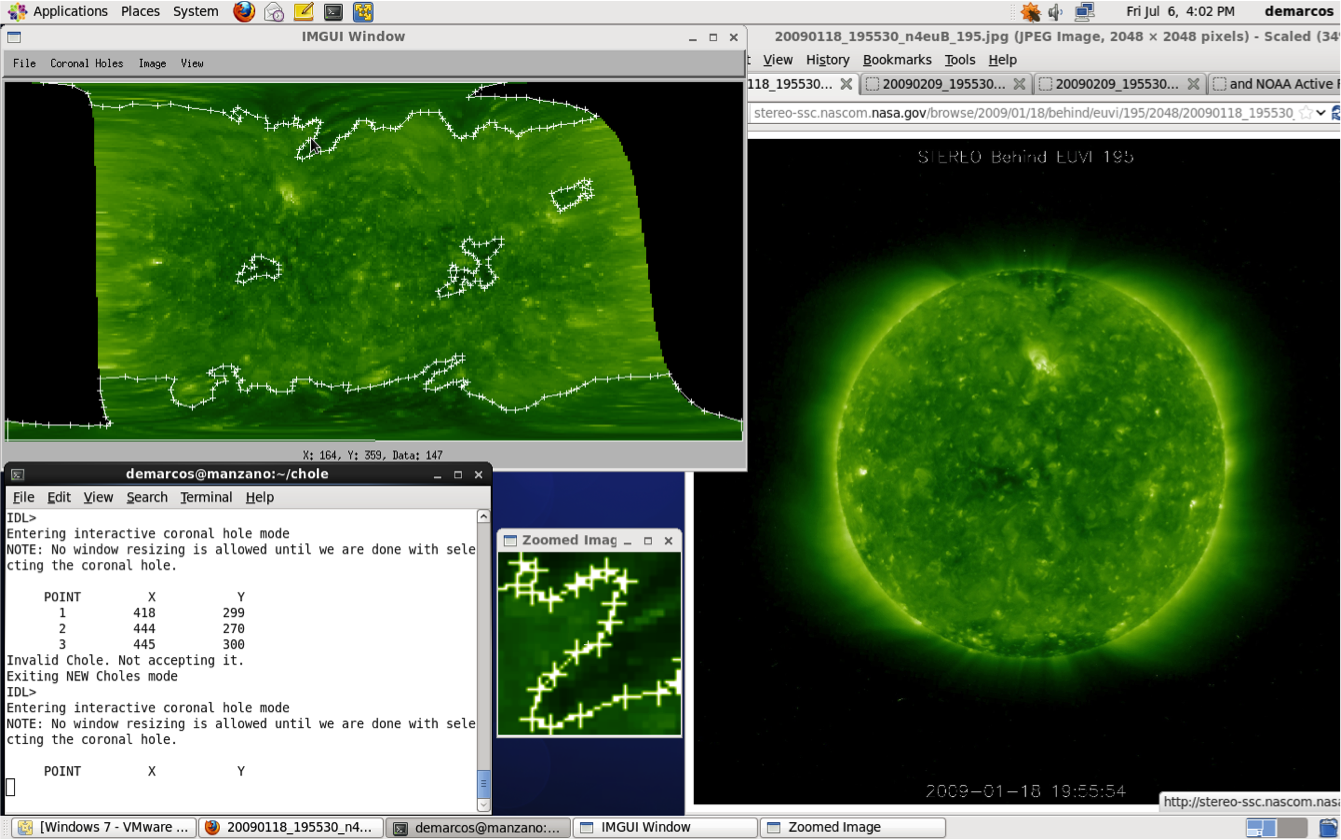}
  \caption{Manual annotation framework developed to segment
    coronal holes.}
  \label{fig:idl_tool}
\end{figure}

Groud truth is necessary in order to assess the performace of coronal hole
detection methods. Unfortunately, the standard practice of comparing
automatic and manual annotations from a single expert is not acceptable
since such an approach can lead to significant biases. A manual
annotation tool developed by Dr.\ Pattichis in the Summer of 2012
has been used to allow multiple users to manually outline the locations
and regions of coronal holes within specific dates. The system was used
to generate coronal hole maps by two different users, and then again to
derive  consensus annotations by combining the two independent maps.
The consensus maps have the unique advantage that they avoid biases
associated with individual raters. Two Carrington rotations are segmented
giving $50$ consensus maps. Fig. \ref{fig:idl_tool} shows the IDL framework
developed to segment coronal holes. .

\section{Coronal hole segmentation}
Coronal holes \cite{altschuler1972coronal} have properties detectable using EUVI
and photo maps.
They appear darker in EUV (Extreme Ultra Violet) spectrum, unipolar,
and do not cross magnetic neutral lines \cite{antiochos2007structure}. Based
on coronal hole properties methods like \cite{Henney2005,krista2009automated}
are developed to segment coronal holes automatically. In
Section~\ref{sec:HHM} we describe Henney Harvey method \cite{Henney2005},
an earlier method that provides a careful integraiton of unipolarity and
darker region requirements.\\

\subsection{ Henney Harvey method}\label{sec:HHM}
The algorithm takes the EUVI and magnetic (photomap) images as
inputs and returns segmented coronal hole image. The input images
are resized to ensure that they share the same resolution.\\
An initial segmentation is estimated on the EUVI image. First,
the dark regions in the EUVI images are detected using spatially adaptive
thresholding (see Fig.~\ref{Fig:HHM algorithm}). The need for adaptive thresholding comes
from the fact that the EUVI images map the spherical surface of the sun to
a rectangular image with pixels of variable size. These dark regions are 
denoised using a morphological {\tt open-close}, followed by an area open
that removes small blobs.\\
Magnetic constraints are then imposed on each estimated EUVI component based
on the corresponding regions in the photomap (magnetic) image. To this end,
the algorithm  needs to determine the polarity of each EUVI component. An
averaging filter is first applied to the polarity image to reduce the noise
level. The polarity of each image is then computed as a ratio of the dominant
polarity to the total number of pixels. The unipolar assumption is enforced
by removing blobs for which the polarity ratio does not exceed 65\%. The remaining
blobs form the estimated coronal hole image.

\begin{figure}
  \begin{algorithmic}[0]
    \Function{HenneyHarvey}{{\tt EUVI}, {\tt mag}}\\
    % VJ: Can you give an equation for the threshold. That would be best here.
    % Use variable names that make sense.
    % Here, I am not using functions. Just pseudocode. You can stay with your
    % functions since you have more space. However, make the code compact!
    ~\\
    \COMMENT{Threshold with area correction.} \\
    \COMMENT{$\theta$ comes from spherical coordinates.} \\
    \COMMENT{$\mu, \sigma$ are estimated over local windows.} \\
    $T$ $\gets$ $\mu - \sigma(0.7 + 0.1 \cos(\theta))$ \\
    \noindent{\tt init\_dark\_img} $\gets$ ({\tt EUVI} $< T$)\\
    ~\\
    \COMMENT{Reduce noise in EUVI image}\\
    \noindent{\tt den\_img} $\gets$ \textbf{open-close} {\tt init\_dark\_img} with {\tt SE}\\
    \noindent{\tt den\_img2} $\gets$ {\tt{\bf remove}} small blobs from {\tt den\_img} \\
    \hspace{1.3 true in}  with {\tt area}$<25$\\
    ~\\
    \COMMENT{Denoise and prepare polarity image}\\
    \noindent{\tt blurred}$\gets$\textbf{blur}  {\tt mag}. \\
    \noindent{\tt pol\_img}$\gets$\textbf{compute\_blob\_polarity} of {\tt blurred}\\
    ~\\
    \COMMENT{Keep unipolar blobs only} \\
    \noindent{\tt coronal\_hole\_img} $\gets$ {\tt{\bf remove}} non-unipolar\\
    \hspace{1.5 true in} blobs from {\tt den\_img2} \\
    \hspace{1.3 true in} with $<65$\% {\tt in pol\_img} \\
    ~\\
    {\textbf return} {\tt coronal\_hole\_img}
    \EndFunction
  \end{algorithmic}
  \caption{Henney-Harvey method for detecting coronal holes. }\label{Fig:HHM algorithm}
\end{figure}

\section{Selecting solar models by matching coronal holes}
The thesis extends prior, initial work by Andrew Delgado where he
defined and solved the problem of automatically selecting physical
models based on observations. The best physical models are selected
by matching the coronal holes between model and consensus. For each
one of the 12 physical models, an optimal physical model is manually
selected for validating the approach. Over 50 maps, the results
indicated that there is a 52\% agreement between the maps that are
visually selected and the ones selected by the automated approach.
A 60\% accuracy is acheived on removing 9 maps for which visual
matching was of low-confidence.\\
The report does not work with coronal hole polarity. This might have
resulted in incorrect clustering and matching of coronal holes having
different polarities. In addition, no clear protocol
was followed when manually selecting the best model. This thesis
addressess these issues by carefully working with polarity,
and following protocol for manually selecting the best model.
% ??? Can we add the following background?
% Levelsets background
% Bi-partite matching with reference to Totally unimodular

% 
% Chapter 3
% 
\chapter{Manual classification of physical models}
In this chapter, we develop a system to support manual classification of physical models.
The problem is introduced in section \ref{sec:subjective_classificaiton_problem}.
Section \ref{sec:ch3mapclassification_solution} describes the manual protocol.

% -------------------------------------------------------------------------------%
% Section: Problem introduction                         %
% -------------------------------------------------------------------------------%
\section{\label{sec:subjective_classificaiton_problem} Introduction}

\begin{figure}[!ht]
  \centering
  \subfloat[Consensus image, size=$720\times360$]
  {
    \includegraphics[width=0.45\textwidth]{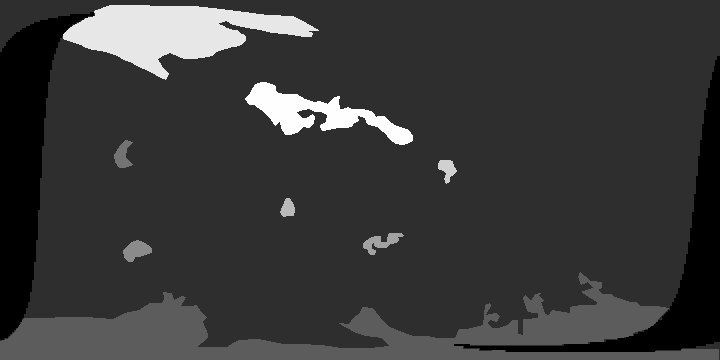}
    \label{subfig:unprocessedConImg}
  }
  \subfloat[Model 1 prediction image, size=$144\times72$]
  {
    \includegraphics[width=0.45\textwidth]{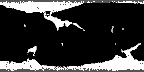}
    \label{subfig:unprocessedM1Img}
  }

  \subfloat[Model 3 prediction image, size=$144\times72$]
  {
    \includegraphics[width=0.45\textwidth]{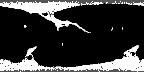}
    \label{subfig:unprocessedM3Img}
  }
  \subfloat[Model 11 prediction image, size=$144\times72$]
  {
    \includegraphics[width=0.45\textwidth]{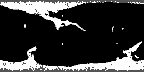}
    \label{subfig:unprocessedM11Img}
  }
  \caption{The physical model classification problem.}
  \label{fig:unprocessedImgs}
\end{figure}

The basic problem of classifying physical models is depicted in Fig. \ref{fig:unprocessedImgs}.
The basic idea is to compare physical model against the consensus maps
and then select the physical models that are closest to the consensus map.
Physically, coronal hole matching requires that we compare clusters of coronal holes
that share the same polarity, similar location, and similar area. 
To appreciate the problem, we note that we have:
\begin{enumerate}
\item \textbf{Matching of many to many:} To see the problem, note that 
  the consensus map of Fig.~\ref{fig:unprocessedImgs} has 8 coronal holes,
  compared to 6 for model 3 (see Fig.~\ref{subfig:unprocessedM3Img}), and
  11 for model 11 (see \ref{subfig:unprocessedM11Img}).

\item \textbf{Classification consistency:} Many physical models appear similar.
  Thus, physical models need to be grouped together based on their similarities.
  Then, instead of classifying individual physical maps, we need to classify
  entire groups.

\item \textbf{Missing observation regions:} The darkest regions of the consensus
  maps represent regions of no observations. Matching needs to take into account that
  we cannot match coronal holes in these regions.
  
\item \textbf{Projection effects} The areas of polar coronal holes are greatly
  exaggerated due to projection effects. To avoid such issues, we will not consider
  visible (manual) matching of regions that are below 30 and above 150 degrees.     
\end{enumerate}

% -------------------------------------------------------------------------------%
% Section: Subjective analysis protocols             %
% -------------------------------------------------------------------------------%
\section{Manual coronal hole matching and physical map classification}\label{sec:ch3mapclassification_solution}
\begin{figure}
  \centering
  \subfloat[Annotated consensus image.]
  {
    \includegraphics[width=0.45\textwidth]{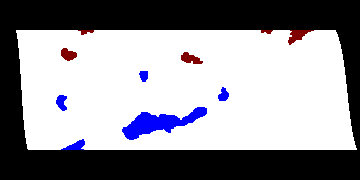}
    \label{subfig:processedConImg}
  }
  \subfloat[Physical model image.]
  {
    \includegraphics[width=0.45\textwidth]{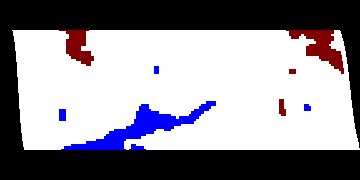}
    \label{subfig:processedM11mg}
  }
  \caption{Visual clustering interface between the consensus map and a physical model image. Positive polarity coronal holes are red. Negative coronal holes are blue. The date is 07/23/2010.}
  \label{fig:preProcImgs}
\end{figure}
This section discusses manual classification protocols designed to address the issues
discussed in Sec~\ref{sec:subjective_classificaiton_problem} while maintaining
reproducibility. Before displaying to user, model and consensus maps are preprocessed to have
the same size. Each coronal hole is color coded based on its polarity (see Fig~\ref{fig:preProcImgs}). Furthermore, we removed regions of no observations, and remove regions with latitude that was below $30^{\circ}$ and above $150^{\circ}$.
The rest of the section describes how to cluster coronal holes, support matching between maps, and the final, manual classification.

\subsection{Manual coronal hole matching using clustering}
In this section, we provide a summary of the coronal hole matching algorithm.
We begin with the clustering rules. 
After clustering, we label each cluster as matchable, new (generated), or
missing (removed).

\begin{figure}[ht!]
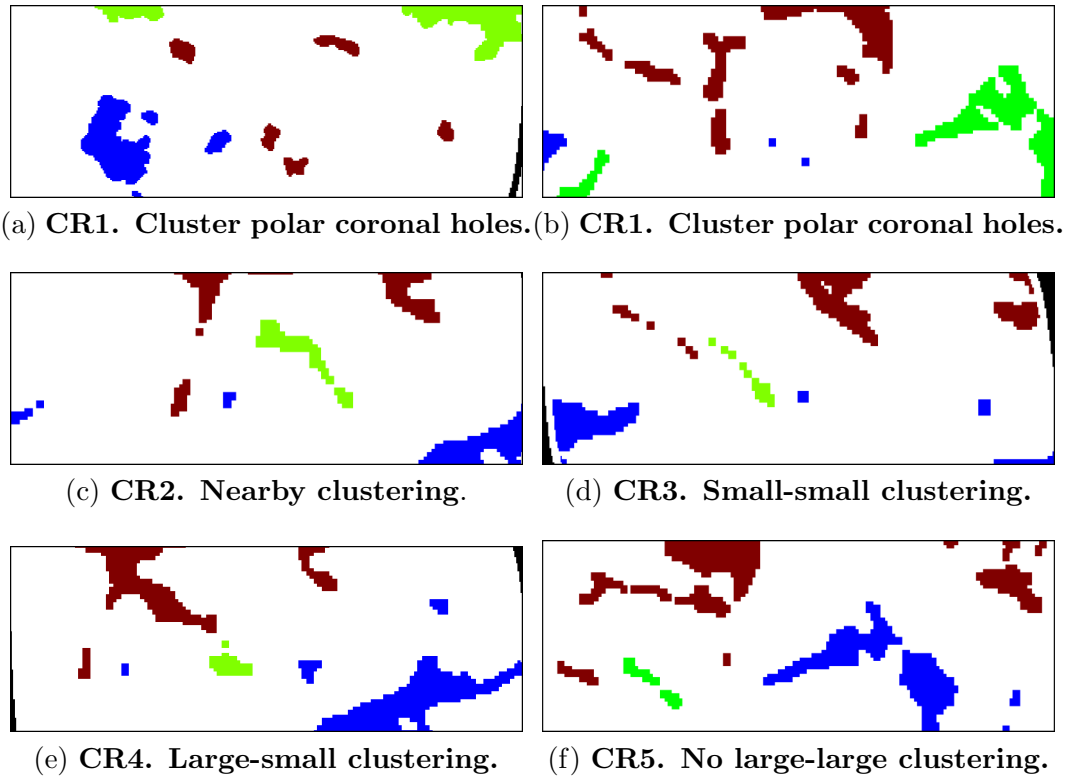

  \subfloat[\textbf{CR1. Cluster polar coronal holes.}]
  {
    \includegraphics[width=0.45\textwidth]{pictures/chapter3/clusteringProtocol/%
      20110125_con_northpole_clustering.png}
    \label{subfig:cr1a}
  }
  \subfloat[\textbf{CR1. Cluster polar coronal holes.}]
  {
    \includegraphics[width=0.45\textwidth]{pictures/chapter3/clusteringProtocol/%
      20110215_m4_southpole_clustering.png}
    \label{subfig:cr1b}
  }

  \subfloat[\textbf{CR2. Nearby clustering}.]
  {
    \includegraphics[width=0.45\textwidth]{pictures/chapter3/clusteringProtocol/%
      20100808_m5_veryclose.png}
    \label{subfig:cr2}
  }
  \subfloat[\textbf{CR3. Small-small clustering.}]
  {
    \includegraphics[width=0.45\textwidth]{pictures/chapter3/clusteringProtocol/%
      20100714_m5_clustersmall.png}
    \label{subfig:cr3}
  }

  \subfloat[\textbf{CR4. Large-small clustering.}]
  {
    \includegraphics[width=0.45\textwidth]{pictures/chapter3/clusteringProtocol/%
      20100803_m9_small_big.png}
    \label{subfig:cr4}
  }
  \subfloat[\textbf{CR5. No large-large clustering.}]
  {
    \includegraphics[width=0.45\textwidth]{pictures/chapter3/clusteringProtocol/%
      20110208_m10_donot_cluster.png}
    \label{subfig:cr5}
  }
  \caption{Coronal holes clustering rules. The rules are applied to the coronal
    holes depicted in green. }
  \label{fig:clusteringInAction}
\end{figure}

For coronal holes of the same polarity, 
we demonstrate the clustering rules in Fig. \ref{fig:clusteringInAction} and 
summarize below:
\begin{description}
\item[\textbf{CR1. Cluster polar coronal holes}:] Coronal holes that were cut 
  at a latitude of $30^{\circ}$ are clustered together into the north polar coronal
  hole cluster. 
  Similarly, coronal holes that were cut at a latitude of $150^{\circ}$ 
  are clustered together into the south polar coronal hole cluster.
\item[\textbf{CR1. Nearby clustering}:] Coronal holes that extremely close to each other
  are clustered together.
\item[\textbf{CR3. Small-small clustering}:] 
  Groups of small coronal holes that are relatively close to each other are
  clustered together.
\item[\textbf{CR4. Large-small clustering}:] 
  A small coronal hole that is close to a much larger one is considered
  part of the larger cluster that involves the larger coronal hole.
  
\item[\textbf{CR5. No large-large clustering}:]
  In general, larger coronal holes are not clustered together
  unless they are extremely close to each other (see \textbf{CR2}).
  \label{des:ch3clustering_rules}
\end{description}

\begin{figure}[ht!]
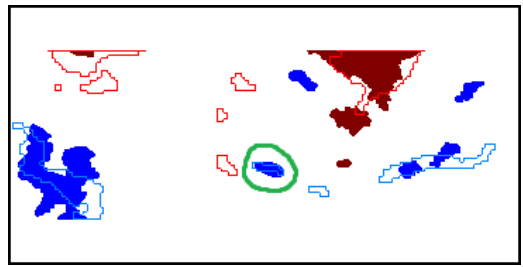
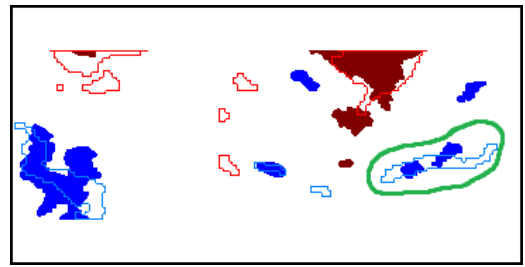
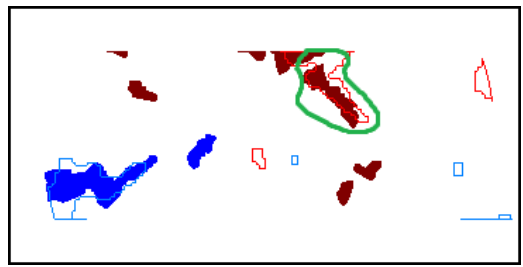
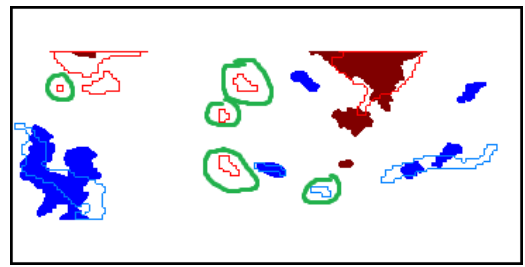
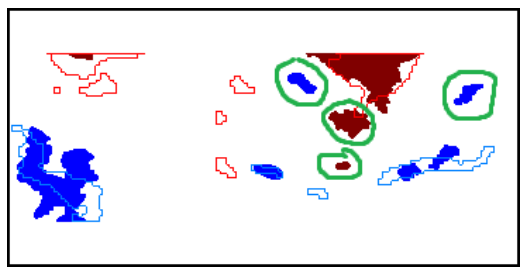

  \centering
  \subfloat[{\bf M1. Polar to polar matching.}]
  {
    \includegraphics[width=0.45\textwidth]{pictures/chapter3/matchingProtocol/%
      20100714_m7_matching.png}
    \label{subfig:PtoP_Similar}
  }
  \subfloat[{\bf M1. Polar to polar matching.}]
  {
    \includegraphics[width=0.45\textwidth]{pictures/chapter3/matchingProtocol/%
      20100714_m9_matching.png}
    \label{subfig:PtoP_Dissimilar}
  }

  \subfloat[{\bf M3. Mid-latitude to mid-latitude matching with good area overlap.}]
  {
    \includegraphics[width=0.45\textwidth]{pictures/chapter3/matchingProtocol/%
      20110120_m1_matching_mtomSimilar.png}
    \label{subfig:MtoM_Similar}
  }
  \subfloat[{\bf M3. Mid-latitude to mid-latitude matching 
    with weak area overlap but good localization.}]
  {
    \includegraphics[width=0.45\textwidth]{pictures/chapter3/matchingProtocol/%
      20110120_m1_matching_mtom_notsimilar.png}
    \label{subfig:MtoM_Dissimilar}
  }

  \subfloat[{\bf M2. Polar to mid-latitude matching.}]
  {
    \includegraphics[width=0.45\textwidth]{pictures/chapter3/matchingProtocol/%
      20100715_m8_matching.png}
    \label{subfig:PtoM}
  }
  \subfloat[Generated (new) coronal holes]
  {
    \includegraphics[width=0.45\textwidth]{pictures/chapter3/matchingProtocol/%
      20110120_m1_matching_gen.png}
    \label{subfig:Gen}
  }

  \subfloat[Removed (missing) coronal holes]
  {
    \includegraphics[width=0.45\textwidth]{pictures/chapter3/matchingProtocol/%
      20110120_m1_matching_rem.png}
    \label{subfig:Rem}
  }
  \caption{\label{fig:matching}
    Coronal hole cluster matching. The regions in green demonstrate the rules.
    Hollow blue regions represent negative coronal hole clusters in the model.
    Hollow red  regions represent positive coronal hole clusters in the model.
    Solid blue  regions represent negative coronal hole clusters in the consensus map.
    Solid blue  regions represent positive coronal hole clusters in the consensus map.}
\end{figure}

Coronal hole clusters of the same polarity are matched based on the following rules 
(see Fig. \ref{fig:matching}):
\begin{description}
\item[{\bf M1. Polar to polar matching:}] Polar clusters with a relatively large area
  overlap (70\% to 100\%) are matched.
\item[{\bf M2. Polar to mid-latitude matching:}]
  A coronal hole cluster from the consensus map that is located
  in the mid-latitude region is matched to a polar cluster from
  the physical model when they overlap by at-least 15\% to 20\%, or more.
\item[{\bf M3. Mid-latitude to mid-latitude matching:}]
  Mid-latitude clusters are matched with good area overlap 
  (e.g., overlap area $>30\%$) or weaker area overlap but
  good localization.
\end{description}
After applying the rules, the remaining coronal hole clusters are
classified as either new (generated in the model)
or removed (missing from the model) (see Fig. \ref{fig:matching}).

\subsection{Map Classification}\label{sec:ch3classificaiton}
To support consistency and reproducibility, maps are pre-classified into two
groups. We use ranks to describe each group. 
In the rank 1 group, we include maps that tend to be closer to the consensus map.
In the rank 2 group, we include maps that tend to be further away from the consensus map.
We then make the final classifications of what
constitutes a good and a bad map based on Fig. \ref{fig:classAlg}.

To decide the rankings, we examine the mid-latitude coronal holes.
Initially, similar to clustering, we group maps based on how similar
they are to each other.
The collection of all of the groups are then classified 
as being closer to the consensus map (rank 1)
or further from the consensus map (rank 2).
Here, we classify a group as being closer to the consensus map if
it contains a substantial number of matched, 
fewer cases of new (generated) and missing (removed) coronal holes.
A ranked group of maps (rank 1 or 2) is then classified a \textit{good match}
if it is in good agreement of the consensus map, where we also
allow slight over-estimation of the area of the coronal holes.
A group of maps that is not considered a \textit{good match} is
classified as a \textit{bad match}. 

\begin{figure}
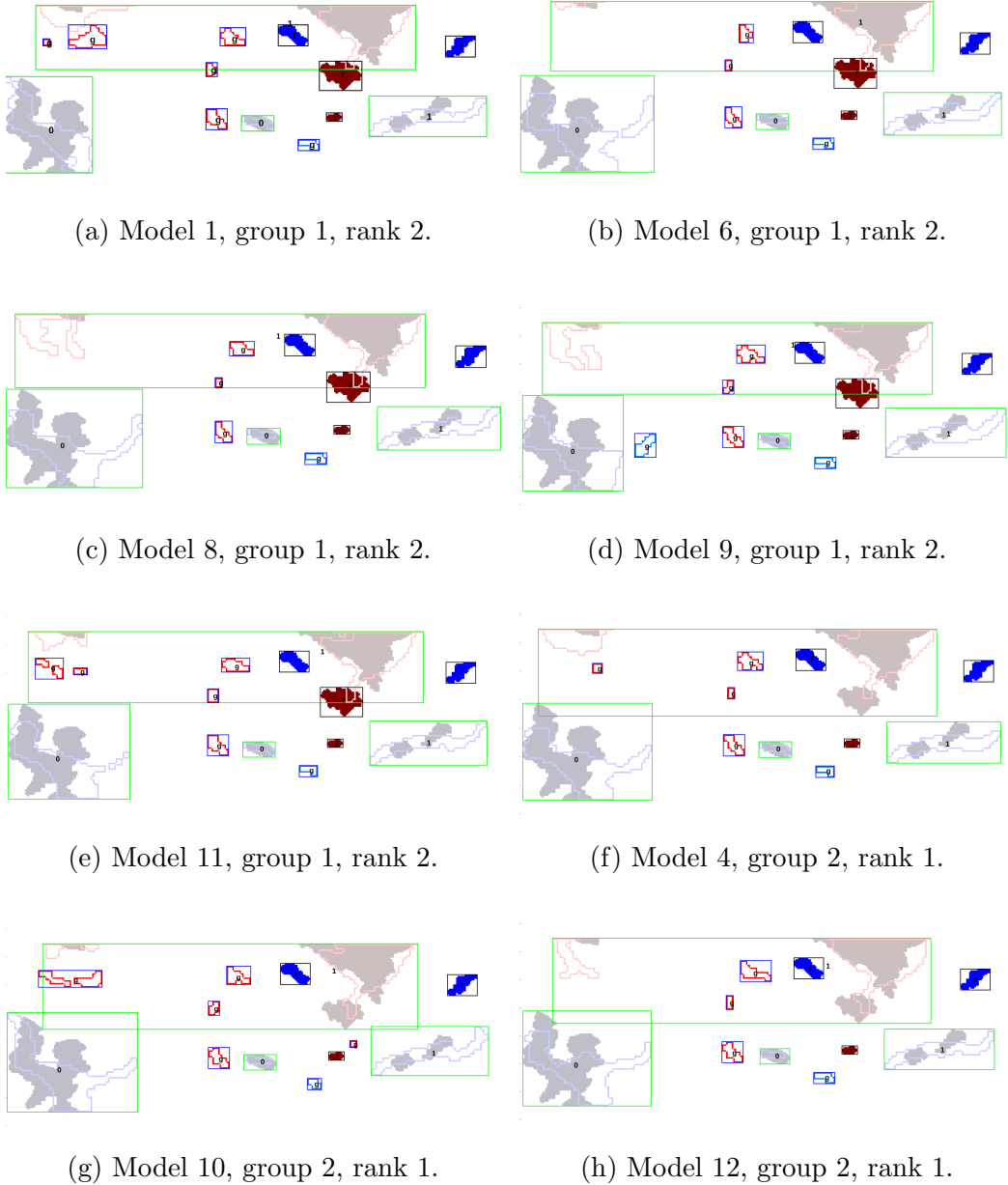

  \subfloat[Model 1, group 1, rank $2$.]
  {
    \includegraphics[width=0.45\textwidth]{pictures/chapter3/classificationProtocol/%
      20110120model1Matched_cropped.png}
    \label{subfig:m1g1}
  }
  \subfloat[Model 6, group 1, rank $2$.]
  {
    \includegraphics[width=0.45\textwidth]{pictures/chapter3/classificationProtocol/%
      20110120model6Matched_cropped.png}
    \label{subfig:m6g1}

  }

  \subfloat[Model 8, group 1, rank $2$.]
  {
    \includegraphics[width=0.45\textwidth]{pictures/chapter3/classificationProtocol/%
      20110120model8Matched_cropped.png}
    \label{subfig:m8g1}

  }
  \subfloat[Model 9, group 1, rank $2$.]
  {
    \includegraphics[width=0.45\textwidth]{pictures/chapter3/classificationProtocol/%
      20110120model9Matched_cropped.png}
    \label{subfig:m9g1}

  }

  \subfloat[Model 11, group 1, rank $2$.]
  {
    \includegraphics[width=0.45\textwidth]{pictures/chapter3/classificationProtocol/%
      20110120model11Matched_cropped.png}
    \label{subfig:m11g1}

  }
  \subfloat[Model 4, group 2, rank $1$.]
  {
    \includegraphics[width=0.45\textwidth]{pictures/chapter3/classificationProtocol/%
      20110120model4Matched_cropped.png}
    \label{subfig:m4g2}

  }

  \subfloat[Model 10, group 2, rank $1$.]
  {
    \includegraphics[width=0.45\textwidth]{pictures/chapter3/classificationProtocol/%
      20110120model10Matched_cropped.png}
    \label{subfig:m10g2}
  }
  \subfloat[Model 12, group 2, rank $1$.]
  {
    \includegraphics[width=0.45\textwidth]{pictures/chapter3/classificationProtocol/%
      20110120model12Matched_cropped.png}
    \label{subfig:m12g2}

  }
  \caption{Map classification example. We show 8 of the 12 physical maps.
    Initially, the maps are grouped into two groups (1 and 2).
    Each group is then assigned a rank (1 or 2).
    The results refer to the maps associated with
    1/20/2011. }
  \label{fig:classificationInAction}
\end{figure}

We present a classification example in Fig. \ref{fig:classificationInAction}.
We begin by explaining the interface.
The filled regions represent coronal hole clusters in the consensus map.
The hollow regions represent coronal hole clusters in the physical map.
In the interface, a rectangular region (not shown here)
was used to specify a matching.
Once the matching has been specified, 
all coronal hole clusters that overlapped
with the user-specified rectangular region
are selected as a match.
A green bounding
box is drawn around the matched coronal hole clusters.
Note that since the bounding boxes are extended out 
to the full size of each coronal hole cluster,
they also contain unmatched coronal holes.
In fact, matched coronal holes are represented 
with faded colors (not bright blue or red).
The remaining coronal hole clusters are automatically
classified as new (generated) or removed (missing).                
They are represented using bright red (positive) or bright blue (negative).

We show examples from two groups in Fig. \ref{fig:classificationInAction}.
Group 1 maps do not have a matching for the positive polarity coronal hole located 
in the upper-right region (depicted as bright red).
Group 2 maps do have a mathing cluster for the same coronal hole
(depicted as faded red).
Furthermore, group 2 maps missed (removed) fewer coronal holes
(depicted as solid blue here).
Thus, group 2 maps are thus classified as rank 1 and group 1 maps 
are classified as rank 2.
Furthermore, rank 1 maps were classified as \textit{good matchings}
and rank 2 maps were classified as \textit{bad matchings}.

\begin{figure}
  \begin{algorithmic}[1]    
    \State \textit{\textbf{Group}} maps based on matched clusters
    \State \textit{\textbf{Rank}} groups as 1 (better) or 2 (worse).      
    \Statex 
    \If {(both ranks represent good matches)}
    \State \textit{\textbf{Classify}} all maps of the day as good matches.
    \EndIf
    \Statex
    \If {(both ranks represent bad matches)}
    \State \textit{\textbf{Classify}} all maps of the day as bad matches.
    \EndIf 
    \Statex
    \If {(rank 1 is acceptable and not rank 2)}
    \State \textit{\textbf{Classify}} maps of rank 1 as good matches.
    \State \textit{\textbf{Classify}} maps of rank 2 as bad matches.
    \EndIf      
  \end{algorithmic}
  \caption{Physical map classification based on group ranking.}
  \label{fig:classAlg}
\end{figure}

% 
% Chapter 4
% 
\chapter{Computer segmentation and classification of physical models}
% Overview
% 
%
\section{Overview}

\section{\label{sec:LSM} Level set method}
We develop a new segmentation method that is based
on the Distance Regularized Level Set Evolution (DRLSE)
method described in \cite{chunming2010}.
DRLSE defines:
\begin{align*}
  d_p(s) &\overset{\vartriangle}{=} p'(s)/s \quad \text{is a divergence operator},  \\
  g     &=1/(1+|\nabla G*I |^2) \quad \text{is the edge function, and} 
  \\
  \delta_\epsilon(x) &\quad\text{is zero for $|x|>\epsilon$ and non-zero for $|x|<\epsilon$}
\end{align*}
where
$p(.)$ is used for defining a regularized distance for the level set function ($\phi$),
$g(.)$ should be minimized at image edges,
$\nabla G$ denotes the gradient of the input image that is computed using
convolution with a derivative of a Gaussian.
The segmented image is computed by evolving the level set as given by:
\begin{align}
  \frac{\partial \phi}{\partial t} &= \mu {\cal R}_p (\phi) 
                                     + \lambda {\cal L}_g (\phi) 
                                     + \alpha    {\cal A}_g (\phi)
                                     \label{eq:leveldef1}
                                     \intertext{where:}
                                     {\cal R}_p (\phi) &= \rm{div}(\rm{d_p}(|\nabla\phi|)\nabla\phi) \quad
                                                         \text{is the distance term,} \nonumber \\
  {\cal L}_g (\phi) &=  \delta_\epsilon(\phi)\rm{div}(g\frac{\nabla\phi}{|\nabla\phi|}), \quad
                      \text{is the boundary term, and} \nonumber \\
  {\cal A}_g (\phi) &= g \delta_\epsilon(\phi) 
                      \quad\text{is an area term.} \nonumber 
\end{align}

We provide a description of the proposed segmentation algorithm in Figs. \ref{fig:mainAlgo} and
\ref{fig:LevelSets}.
The approach requires 
joint processing of the EUVI and magnetic images.
Most importantly, we need
to modify the edge function so that it does not allow crossing the magnetic neutral lines.
This is accomplished by modifying the edge function to be:
\begin{equation}
  {\tt pg = (1-p) g }
\end{equation}
where ${\tt p}$ assumes the value of 1 over the magnetic polarity 
boundaries detected in the
magnetic image and is zero away from the boundary (see Fig. \ref{fig:mainAlgo}). 
Thus, over the magnetic lines, the edge function becomes zero and prevents crossing of the neutral line boundary.

From \eqref{eq:leveldef1}, we have found that $\alpha$ and
the spatial spread of the Gaussian ($\sigma$) used for computing 
the edge function are the two parameters that can affect overall segmentation
performance.
To find the optimal parameter values, we compare against
the consensus maps, and look for the optimal values
using (e.g., see \cite{Pattichis2014}):
\begin{equation}
  \min_{\alpha, \, \sigma} \sqrt{[1-{\tt spec}(\alpha, \sigma)]^2 + [1-{\tt sens}(\alpha, \sigma)]^2} 
  \label{eq:basicOpt}
\end{equation}
where ${\tt Spec}$ denotes the (pixel-level) specificity and ${\tt Sens}$ denotes the 
corresponding sensitivity.
The solution of \eqref{eq:basicOpt} gives the optimal values for each image.
For each image, we constrain the optimization problem \cite{6693707, rodriguez2014system, 6189990,murillo2010global, Esakki2016} for 
$\alpha\in[-3, +3], \, \sigma\in [0.2, 1]$.
Over the training set, we select the median values
over the entire set. 
We then report the performance over the testing set.
In the results section, we report the performance
of the algorithm using leave-one-out.

The optimization of \eqref{eq:basicOpt} is challenging since
derivative estimates can be very noisy.
To this end, we use a robust optimization method
based on {\tt Pattern-search} initialized with $\alpha_0=0, \sigma_0=0.5$.
We refer to \cite{optbook} for details on the optimization procedure.       
Furthermore, to speed-up convergence, we initialize the segmentation algorithm using the Henney-Harvey algorithm
as documented in Fig. \ref{fig:mainAlgo}.

\begin{figure}
  \begin{algorithmic}[0]
    \Function{Segment}{{\tt EUVI}, {\tt mag}, $\alpha$, $\sigma$}\\
    \noindent$I \gets$ \textbf{smooth} {\tt syn\_img} with $15\times 15$ Gaussian kernel \\
    \noindent\hspace{0.4 true in}with optimization variable $\sigma$.\\
    {\tt g}   $\gets$ $\frac{1}{1+(I_x^2+I_y^2)}$ \\
    ~\\
    \COMMENT{Make $g$ zero at magnetic boundaries} \\
    {\tt p}   $\gets$ \textbf{DetectMagneticCrossLines}({\tt mag\_img})\\
    {\tt pg} $\gets$ (1 - {\tt p} ) .* {\tt g} \\
    ~\\
    \COMMENT{Initialize with HenneyHarvey segmentation method}\\
    \noindent$\phi_0 \gets$ \textbf{HenneyHarvey} ({\tt syn\_img}, {\tt photo\_img})\\
    ~\\
    \COMMENT{Run with modified edge function}\\
    \COMMENT{ and optimization parameter $\alpha$.}\\
    \noindent{\bf return} {\tt LS}({\tt EUVI}, {\tt mag}, $\phi_0$, pg, $\alpha$)
    \EndFunction
  \end{algorithmic}
  \caption{Main coronal hole segmentation algorithm. 
    The parameters $\alpha$ and $\sigma$ are optimized using {\tt Pattern-Search}.
  }\label{fig:mainAlgo}
\end{figure}

\begin{figure}
  \begin{algorithmic}[0]
    \Function{LS}{{\tt EUVI}, {\tt mag}, {\tt $\phi_0$, {\tt pg}, $\alpha$}}
    \State $\phi$ $\gets$ $\phi_0, \quad$ \COMMENT{init. using previous method}
    \For{i $\leq$ n }
    \State $\delta$($\phi$) $\gets$ \textbf{Dirac}($\phi$, $\epsilon$) \\
    ~\\
    \hspace{0.5 true in}\COMMENT{Use modified edge function {\tt pg}:}
    \State $F_a$ $\gets$ \textbf{areaTerm}($\delta$($\phi$), {\tt pg}) \
    \State $F_e$ $\gets$ \textbf{edgeTerm}($\delta$($\phi$), $\phi$, {\tt pg})
    \State $F_d$ $\gets$ \textbf{Regularize\_distance}($\phi$)\\
    ~\\
    \hspace{0.5 true in}\COMMENT{Allow $\alpha$ to vary for optimization:}
    \State $\phi$ $\gets$ $\phi$ + ts$\cdot$($\mu F_d$ + $\lambda F_e$ + $\alpha F_a$)
    \EndFor \\
    % ???VJ: I think you need phi=0. Not <0. Please check! \\
    \noindent{\bf return} $\delta(\phi)$
    \EndFunction
  \end{algorithmic}
  \caption{Level-set segmentation algorithm using
    the modified edge function {\tt pg} and $\alpha$.}\label{fig:LevelSets}
\end{figure}

\section{Classification system overview}
An outline of the algorithm is provided in Fig. \ref{Fig:autoClassOverviewAlg}.
The algorithm accepts the set of physical maps and a reference map that corresponds
to the physical observations.
Here, for the best results, the reference map should be the consensus map.
Alternatively, we can set the reference map to a manual or an automated segmentation
map and measure its performance against the consensus map.

\begin{figure}[h!]
  \begin{algorithmic}[1]
    \Function{map\_classification}{{\tt dates}}\\
    \COMMENT{\textbf{Input:} {\tt dates} to process.}\\
    \COMMENT{Classify physical maps for given {\tt dates}.}\\
    \State~\C{Process each date separately}
    \For{ {\tt date} $\in$ {\tt dates}}
    \State \C{Read  and process reference image}
    \State \C{ (e.g., Consensus image, or automatically segmented images).}
    \State {\tt ref\_map} $\gets$ {\bf load\_ref\_data}({\tt date})
    \State {\tt ref\_map$_{\{+,-\}}$} $\gets$ {\bf pre\_process}({\tt ref\_map})\label{alg:ov_preprocessing1}
    ~\\
    \State~\C{Process associated physical models}
    \For{ {\tt model} $\in \{{\tt model\_1}, \dots, {\tt model\_12}\}$ }
    \State {\tt model\_map} $\gets$ {\bf load\_model}({\tt date}, {\tt model})
    \State {\tt model\_polarity\_map$_{\{+,-\}}$} $\gets$ {\bf pre\_process}({\tt mod\_img})\label{alg:ov_preprocessing2}
    ~\\
    \State~\C{Analyze each polarity separately}
    \For{polarity {\tt p} $\in$ \{ $+$, $-$\}}
    \State {\small \textbf{\textit{Cluster}}} coronal holes that are are very close.
    \State {\small \textbf{\textit{Detect}}}  coronal hole clusters that are in
    \State $\qquad$ physical maps but not in reference map
    \State $\qquad$ using Mahalanobis distance threshold and
    \State $\qquad$ store the results in {\tt new\_map$_p$} and {\tt missing\_map$_p$}
    \State {\small \textbf{\textit{Re-cluster}}} remaining coronal holes in {\tt ref\_map} and
    \State $\qquad$  {\tt model\_map} to achieve equal number of clusters.
    \State {\small \textbf{\textit{Match}}} clusters using \textit{linear programming} 
    \State $\qquad$ and save the results in {\tt matched\_map$_p$}                                                            
    \EndFor
    \State {\small \textbf{\textit{Extract}}} {\tt features}  from {\tt new\_map$_p$}, {\tt missing\_map$_p$}
    \State $\qquad$ and {\tt matched\_map$_p$} for polarity {\tt p} $\in$ \{+, -\}.         
    \State {\small \textbf{\textit{Classify}}} {\tt model} using extracted features.
    \label{algLine:get_features}
    \EndFor
    \EndFor 
    \EndFunction
  \end{algorithmic}
  \caption{Overview of physical map classification algorithm.}
  \label{Fig:autoClassOverviewAlg}
\end{figure}

The main algorithm accepts the dates that need to be processed.
Then, for each date, we load the physical models and the reference map.
All of the maps are pre-processed prior to matching.
Very near coronal holes are clustered and a comparison between each 
physical model and the reference map reveals new and missing coronal
hole clusters that are stored in new maps, according to their polarity.
The remaining coronal hole clusters are matched using Linear Programming.
Features extracted from all maps then used for the final classification.

The rest of the chapter is organized into four subsections.
In section \ref{sec:ch4preprocessing}, we describe the pre-processing steps.
Section \ref{sec:ch4clustering_close} describes clustering of very close
coronal holes. Section \ref{sec:ch4detecting_genrem} describes the process for detecting new and 
missing coronal hole clusters. The matching process which involves re-clustering
is described in section \ref{sec:ch4Matching}. The classification step is described in section \ref{sec:ch4classificaiton}.

\section{Map preprocessing}\label{sec:ch4preprocessing}
The pre-processing steps are listed in Fig. \ref{fig:preProcessing}.
The basic steps involve resizing the maps to the same resolution,
removing regions near the poles and regions where we have no observations, and
splitting the maps into positive and negative polarity maps.

\begin{figure}[ht!]
  \begin{algorithmic}[1]  
    \Function{pre\_process}{{\tt maps}}
    \State \COMMENT{Pre-process maps to remove regions of no observations, polar regions, and} 
    \State \COMMENT{ split them based on polarity.}
    ~\\
    \State \COMMENT{Extract reference and model maps}
    \State [{\tt ref\_map}, {\tt model\_map}] $\gets$ {\small \textbf{extract\_coronal\_hole\_maps}} ({\tt maps})
    \State
    \State \COMMENT{Remove small gaps in the maps}
    \State [{\tt ref\_map}, {\tt model\_map}] $\gets$ {\bf binary\_close~}({\tt ref\_map}, {\tt model\_map})
    ~\\
    \State \COMMENT{Resize maps to photomap image size}
    \State [{\tt ref\_map}, {\tt model\_map}] $\gets$ {\bf resize\_to\_same\_size~}({\tt ref\_map}, {\tt model\_map})
    ~\\
    \State \COMMENT{Extract regions based on polarity in magnetic images}
    \State [{\tt ref\_magnetic}, {\tt model\_magnetic}] $\gets$ {\small \textbf{extract\_magnetic\_maps}} ({\tt maps})
    
    \State {\tt ref\_map$_{+,-}$} $\gets$ {\bf extract\_polarity\_maps~}({\tt ref\_map}, {\tt ref\_magnetic})
    \State {\tt model\_map$_{+,-}$} $\gets$ {\bf split\_coronal\_maps~}({\tt model\_map}, {\tt model\_magnetic})
    ~\\
    \State \COMMENT{Remove regions where there are no observations,}
    \State \COMMENT{ latitude 0 to 30 degrees, and 150 to 180 degrees}
    \State {\tt no\_data\_map} $\gets$  {\bf set\_no\_obs\_regions~}({\tt maps})
    ~\\
    \State \COMMENT{Remove no data regions from all maps}
    \State {\tt ref\_map$_{+,-}$}  $\gets$  {\bf remove\_no\_data} ({\tt ref\_map$_{+-}$},  {\tt no\_data\_map})
    \State {\tt model\_map$_{+,-}$} $\gets$ {\bf remove\_no\_data} ({\tt mod\_map$_{+,1}$}, {\tt no\_data\_map})
    \EndFunction
  \end{algorithmic}  
  \caption{\label{fig:preProcessing}Pre-processing reference maps and model maps prior to analysis.}
\end{figure}

% ----------------------------------------------------------------------------------%
% Clustering near by coronal holes.
% -----------------------------------------------------------------------------------%
\section{Clustering very close coronal holes.}\label{sec:ch4clustering_close}
Model maps could contain coronal holes that are very close. Processing these coronal
holes individually may lead to improper working of \textbf{\textit{Detection}} and
\textbf{\textit{Matching}} algorithms. Hence we cluster very near coronal holes. Clustering
is based on minimum pixel distance between coronal holes. The basic idea
is to cluster coronal holes iteratively till there are no more coronal holes which
are seperated by a threshold.

% ----------------------------------------------------------------------------------%
% Detecting generated and removed coronal holes
% ----------------------------------------------------------------------------------%
\section{Detecting new and missed coronal hole clusters}\label{sec:ch4detecting_genrem}
A coronal hole cluster that is present in the reference map may be \textit{missing} from the 
physical model map.
Alternatively, a physical model may have \textit{new} coronal hole clusters that are absent
from the reference map.
Clearly, \textit{new} and \textit{missing} coronal hole clusters are to be removed
from Both maps, prior to matching.   
In this section, we describe how to detect them.

To detect \textit{new} coronal hole clusters, we examine each coronal hole cluster in the physical model and
find the nearest corresponding one in the reference map. Then, we compute the Mahalanobis distance between the origin (perfect match) and the vector composed of the minimum physical distance between
the coronal holes and their physical area difference. If the Mahalanobis distance, the coronal hole in the physical model is classified as new.

Similarly, to detect \textit{missing} coronal hole clusters, we examine each coronal hole cluster in the reference map and look for the corresponding one in the physical map.
Then, a coronal hole cluster in the reference map is classified as \textit{missing} if
is has a high Mahalanobis distance from the origin.

\section{Matching with re-clustering}\label{sec:ch4Matching}
After removing the new and missing coronal hole clusters, the 
remaining ones need to be matched. Unfortunately, we
can still have different numbers of clusters in each map.
Thus, instead of matching  maps having different number
of clusters, we need to first combine them
together to have equal numbers of clusters. Clustering
is accomplished using the minimum physical distance between coronal holes.
The basic idea is to iteratively cluster together all
coronal hole clusters that are separated by a minimal
physical distance until we reach the desired number of
clusters. After clustering we introduce linear programming model for
computing an optimal matching between coronal hole clusters.

Let $i$ be used to index clusters in the reference map.
Similarly, let $j$ be used to index clusters in the physical map.
Then, we use  $m_{i,j}$
to denote a possible match between cluster $i$ in the reference map
and cluster $j$ in the physical map.
Thus, $m_{i,j}=1$ when there is a match between the clusters
and $m_{i,j}=0$ otherwise.
We also assign a cost $w_{i,j}$ associated with the matching.
Here, we set the $w_{i,j}$ to be the shortest spherical distance
between the clusters.
Thus, $w_{i,j}=0$ when the clusters overlap.

Formally, we find an optimal matching by solving:
\begin{align}
  & \min_{m_{i,j}} \,\, \sum_i \sum_j w_{i,j} m_{i, j} 
    \label{eq:min_problem} \\
  \intertext{subject to:}          
  & \sum_i m_{i, j} = 1, \label{eq:constri} \\
  & \sum_j m_{i, j} = 1  \label{eq:constrj} \\
  & m_{i, j} \in \{0, 1\} 
\end{align}
where $m_{i, j}$ denotes the assignment that minimizes
the weighted matching of \eqref{eq:min_problem},
while each cluster can only be assigned to one
other cluster as required by \eqref{eq:constri} and \eqref{eq:constrj}.
This is a typical bipartite matching setup, making matching matrix
created from $m_{i,j}$ to be totally unimodular. As discussed in
\cite{papadimitriou1982combinatorial} this problem when solved with
linear programming will return an integer solution.

\section{Classification}\label{sec:ch4classificaiton}
Each physical map is finally classified as {\it good} or {\it bad}
(see section \ref{sec:ch3classificaiton})
% num_gen_max   <- max(cur_df['num_gen_vec'])
% num_rem_max   <- max(cur_df['num_rem_vec'])
% num_gr_max    <- max(cur_df['num_gr_vec' ])
% gen_ar_max    <- max(cur_df['gen_ar_vec' ])
% rem_ar_max    <- max(cur_df['rem_ar_vec' ])
% gr_ar_max     <- max(cur_df['gr_ar_vec'  ])
% mat_ar_max    <- max(cur_df['mat_ar_vec' ])
The following features are used for classification:
\begin{itemize}
\item {\bf Number of new coronal holes}: Number of coronal holes that are predicted by model but absent in reference map.
\item {\bf Number of missing coronal holes:} Number of coronal holes that are missing from model but present in reference map.
\item {\bf Total area of new coronal holes}: Total area of generated coronal holes projected onto an unit sphere.
\item {\bf Total area of missing coronal holes:} Total area of removed coronal holes projected onto an unit sphere
\item {\bf Over estimated area}: Area overestimated by model.
\end{itemize}
% ??? VJ: Consider removing the PCA.
Following feature extraction, we apply principal component analysis to reduce the features.
Classification is performed using k-nearest neighbors (KNN) $K=11$ and SVM.

\begin{comment}
  \begin{figure}
    \begin{algorithmic}[1]
      \State \COMMENT{Write about inputs to funciton}
      \Function{validity\_classifier}{{\tt labels}, {\tt auto\_imgs}, {\tt dates}}
      \State {\tt feat} $\gets$ {\bf extract\_features} ({\tt auto\_imgs})
      \State {\tt pred\_labels} $\gets$ {\bf Empty}(){\COMMENT \blu~Is this correct way to declare empty array}
      \For{{\tt cur\_dt} {\it in} {\tt dates}}
      \State {\tt feat\_tst}   $\gets$ {\bf get\_tst\_features}({\tt feat},    {\tt cur\_dt})
      \State {\tt feat\_trn}   $\gets$ {\bf get\_trn\_features}({\tt feat},    {\tt cur\_dt})
      \State {\tt lab\_tst}    $\gets$ {\bf get\_tst\_labels}  ({\tt labels},  {\tt cur\_dt})
      \State {\tt model}       $\gets$ {\bf build\_classificaiton\_model}({\tt feat\_trn}, {\tt lab\_trn})
      \State {\tt pred\_labels}$\gets$ {\bf append}( pred\_labels, {\bf apply\_model}({\tt model}, {\tt feat\_tst})
      \EndFor
      \State \Return{\tt{pred\_labels}}
      \EndFunction
    \end{algorithmic}
    \caption{Classification algorithm using leave one out cross validation}
    \label{fig:ch4classificaitonAlg}
  \end{figure}
\end{comment}

\chapter{Results}
\section{Summary}
The results are summarized in six sections.
Section \ref{sec:ground_truth} provides a description of the dataset.
Section \ref{sec:LSM_res} provides results for coronal hole segmentation.
Section \ref{sec:clustering} provides clustering examples.
Section \ref{sec:detection} summarizes results for coronal hole detection.
Pre-clustering and matching between physical and consensus maps are given in  \ref{sec:matching}.
Final classification results are given in section \ref{sec:classification_results}.

\section{Classification dataset}\label{sec:ground_truth}
The dataset consisted of two Carrington rotations \cite{howard1981surface}
that consisted of 50 days. The first Carrington rotation covers the dates 
from 07/13/2010 to 08/09/2010. The second Carrington rotation covers the
dates from 01/20/2011 to 02/16/2011. For each day, we have
\begin{itemize}
	\item Synoptic image \cite{altschuler1977high}
	\item Magnetic photo map image
	\item Two human segmentations
	\item Consensus map derived from the two human segmentations
	\item Twelve coronal hole prediction maps
\end{itemize}
Consensus maps are used to manually determine good and
bad prediction maps. For comparing to human performance, 
we also compare the results against the use of
two human segmentations (human raters labeled R4 and R7).
For training, we use leave-one-out over 10 randomly chosen
dates and report results on the remaining 40.

\section{\label{sec:LSM_res} Segmentation}

\begin{figure}[ht]
  \centering
  \subfloat[EUVI]
  {\includegraphics[width=0.4\textwidth]{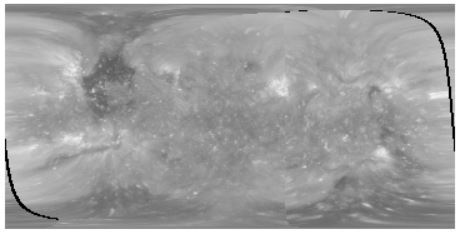}
    \label{fig:subfigure4}} \\
  \subfloat[Consensus map]
  {\includegraphics[width=0.49\textwidth]{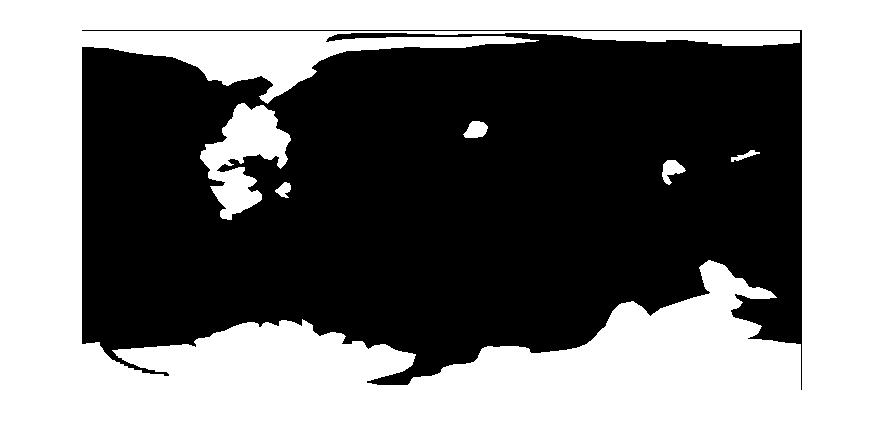}
    \label{fig:subfigure1}} \\
  \subfloat[{\tt Henney-Harvey} Method, ${\tt unit\_dist} = 0.19$]
  {\includegraphics[width=0.4\textwidth]{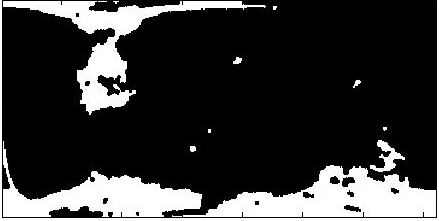}
    \label{fig:subfigure2}} \\
  \subfloat[Level-set segmentation, ${\tt unit\_dist} = 0.09$]
  {\includegraphics[width=0.4\textwidth]{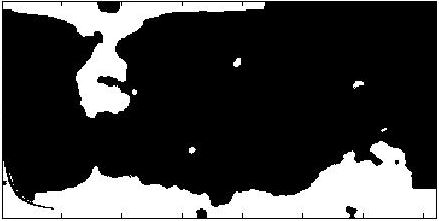}
    \label{fig:subfigure3}} \\
  \caption{Best case results for the level-set segmentation method
    for the input data from January, 24th, 2011.
    We have a reduction in the unit distance by 50.76\%
    (see \eqref{eq:basicOpt} for definition of unit distance).}
  \label{fig:bestcase}
\end{figure}

\begin{figure}[ht]
  \centering
  \subfloat[EUVI]
  {\includegraphics[width=0.4\textwidth]{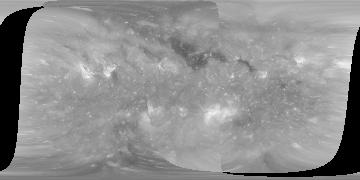}
    \label{fig:subfigure1}} \\
  \subfloat[Consensus map]
  {\includegraphics[width=0.4\textwidth]{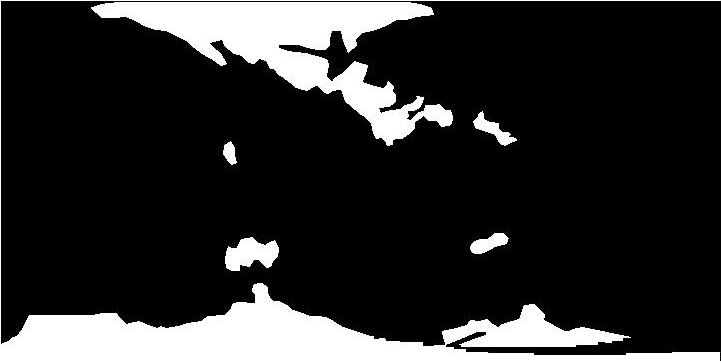}
    \label{fig:subfigure2}} \\
  \subfloat[{\tt Henney-Harvey} Method, ${\tt unit\_dist} = 0.17$]
  {\includegraphics[width=0.4\textwidth]{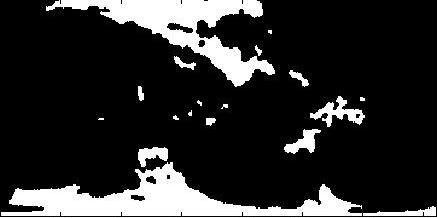}
    \label{fig:subfigure3}} \\
  \subfloat[Level set segmentation, ${\tt unit\_dist} = 0.12$]
  {\includegraphics[width=0.4\textwidth]{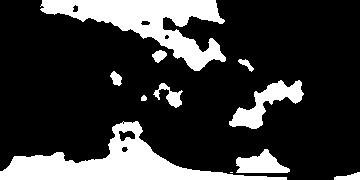}
    \label{fig:subfigure4}}
  \caption{Typical case results for the level-set segmentation method
    for the input data from May 2, 2011.
    We have a reduction in the unit distance by 29.36\%
    (see \eqref{eq:basicOpt} for definition of unit distance).}            
  \label{fig:typicalcase}
\end{figure}

\begin{figure}[ht]
  \centering
  \subfloat[EUVI]
  {\includegraphics[width=0.4\textwidth]{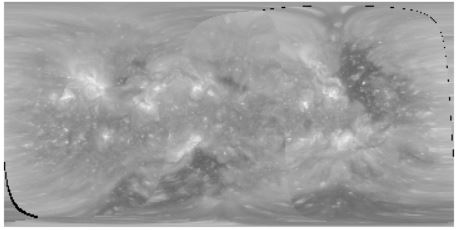}
    \label{fig:subfigure4}}
  \quad
  \subfloat[Consensus map]
  {\includegraphics[width=0.4\textwidth]{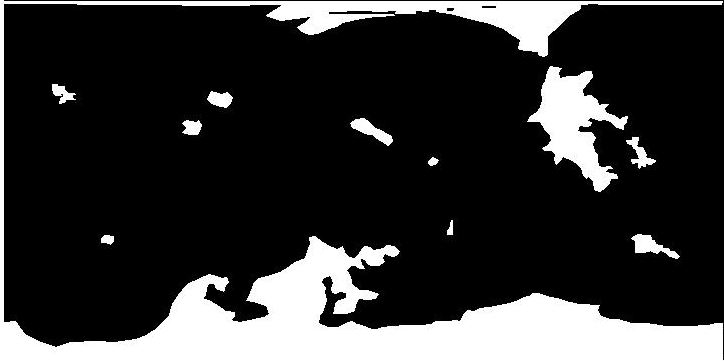}
    \label{fig:subfigure1}}
  \quad
  \subfloat[{\tt Henney-Harvey} method, ${\tt unit\_dist} = 0.14$]
  {\includegraphics[width=0.4\textwidth]{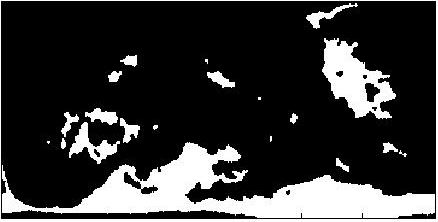}
    \label{fig:subfigure2}}
  \quad
  \subfloat[Level set segmentation, ${\tt unit\_dist} = 0.17$]
  {\includegraphics[width=0.4\textwidth]{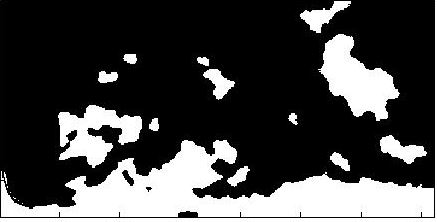}
    \label{fig:subfigure3}}
  \caption{Worst case results for the level-set segmentation method
    for the input data from February 29, 2011.
    In this case, we have an increase in the unit distance by 24.96\% 
    (see \eqref{eq:basicOpt} for definition of unit distance).}
  \label{fig:badcase}
\end{figure}

We present results for the consensus maps ($N=46$) in Table \ref{table:summary}.
On average, we have a reduction of 19\% ($\sigma=17.7\%$) in the unit-distance 
from the ideal segmentation.
On the other hand, we also have cases where the new method performs worse
than the original {\tt Henney-Harvey} method.
We provide three representative examples in Figs. \ref{fig:bestcase},
\ref{fig:typicalcase}, and \ref{fig:badcase}. 

The best case scenario is shown in Fig. \ref{fig:bestcase}.
Here, it is clear that the level-set method provides smoother boundaries
with coronal hole estimates that are better filled and in better
agreement with the {\tt Henney-Harvey} method.
On the other hand, there are significant gaps in the original method.      

Similar comments apply for the typical case shown in Fig. \ref{fig:typicalcase}.
Overall, a larger number of coronal holes appear in the original method that
cannot be found in the consensus map.
On the other hand, most of the smaller coronal holes are missing from the 
level-set approach.

For both methods, the worst case is shown in Fig. \ref{fig:badcase}.
In this example, both methods fail to fully detect the coronal hole in the north pole
region.
A careful examination of the EUVI image of Fig. \ref{fig:badcase}(a) 
shows that the pixels in this north-pole coronal hole are much brighter
than average.
Thus, the initialization by the {\tt Henney-Harvey} method fails to detect
the coronal hole in the south pole and
the level-set evolution does not recover from this initialization.
Furthermore, the consensus map appears to have finer resolution detail
than both maps.
In this case, the level-set approach overly smooths the detected components
as compared to the consensus map and the original method.        

\begin{table}[ht]
  \center
  \caption[caption]{
    Percentage improvement of proposed level-set segmentation method 
    (see \eqref{eq:basicOpt} for definition of unit distance).
    Overall, we have an average (mean) improvement of $\textbf{19.01}\%$
    with a standard deviation of $\textbf{17.7}\%$}
  \label{table:summary}.
  \begin{tabular}{|c|c|c|c|c|}
    \hline \textbf{Order Stat.}  & \textbf{Henney-Harvey}  & \textbf{Level Sets}   & \textbf{\% Impr.} \\
    \hline \hline
    Min                            &       0.14    &       0.17    &       -24.96  \\      \hline
    25\%                   &       0.26    &       0.27    &       -2.00   \\      \hline
    50\%                   &       0.31    &       0.27    &       12.06   \\      \hline
    75\%                   &       0.23    &       0.16    &       30.44   \\      \hline
    Max                            &       0.19    &       0.09    &       50.76   \\      \hline
  \end{tabular}
\end{table}

\section{Clustering}\label{sec:clustering}
\begin{figure}
  \subfloat[Positive model map before clustering]
  { \includegraphics[width=0.47\textwidth]{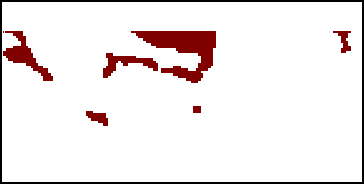}
    \label{subfig:pos_model}}
  \subfloat[Positive model map after clustering closest coronal holes]
  { \includegraphics[width=0.47\textwidth]{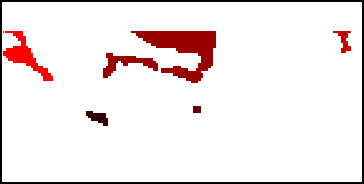}
    \label{subfig:clustered_pos_model}}

  \subfloat[Negative model map before clustering]
  { \includegraphics[width=0.47\textwidth]{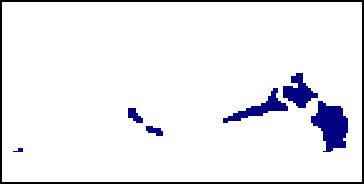}
    \label{subfig:neg_model}}
  \subfloat[Negative model map after clustering closest coronal holes]
  { \includegraphics[width=0.47\textwidth]{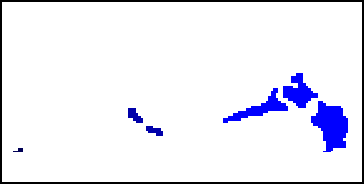}
    \label{subfig:clustered_neg_model}}
  \caption{Clustering model}
  \label{fig:clustering_model_map}
\end{figure}

\begin{figure}
  \subfloat[Positive consensus map before clustering]
  { \includegraphics[width=0.47\textwidth]{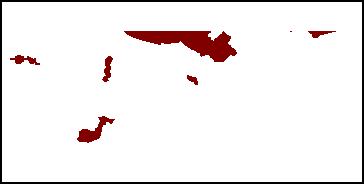}
    \label{subfig:pos_consensus}}
  \subfloat[Positive consensus map after clustering closest coronal holes]
  { \includegraphics[width=0.47\textwidth]{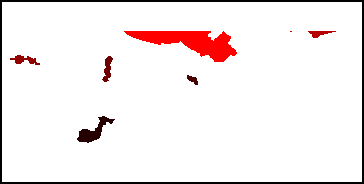}
    \label{subfig:clustered_pos_consensus}}

  \subfloat[Negative consensus map before clustering]
  { \includegraphics[width=0.47\textwidth]{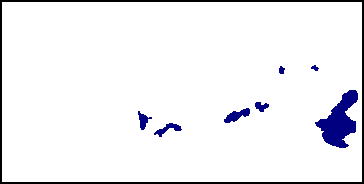}
    \label{subfig:neg_consensus}}
  \subfloat[Negative consensus map after clustering closest coronal holes]
  { \includegraphics[width=0.47\textwidth]{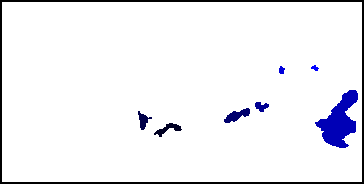}
    \label{subfig:clustered_neg_consensus}}
  \caption{Clustering consensus}
  \label{fig:clustering_consensus_map}
\end{figure}

Coronal holes which are very close to each other are clustered together based
on pixel distance. In this section clustering in reference map and model
map are demonstrated in Fig. \ref{fig:clustering_model_map} and Fig.
\ref{fig:clustering_consensus_map}. Clustered coronal holes are filled
with same shade of red or blue. Shades of red is used to mark
positive clusters while blue marks  negative clusters. As seen
from Fig. \ref{fig:clustering_consensus_map} consensus maps rarely
have very near coronal holes.
            
\section{Detection of new and missing coronal holes}\label{sec:detection}
In order to detect new and missing coronal holes, 
the distance threshold parameter was experimentally set on the ten randomly chosen
dates.
Then, independent testing on the remaining 40 dates was $87.7\%$.

\begin{figure}
  \subfloat[Consensus map]
  { \includegraphics[width=0.47\textwidth]{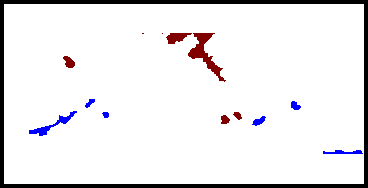}
    \label{subfig:gen_rem_good_consensus}}
  \subfloat[Model map]
  { \includegraphics[width=0.47\textwidth]{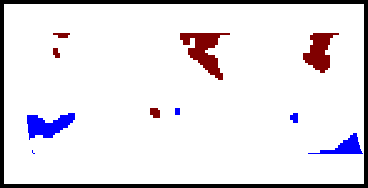}
    \label{subfig:gen_rem_good_model}}

  \subfloat[Coronal holes missing from model map.]
  { \includegraphics[width=0.47\textwidth]{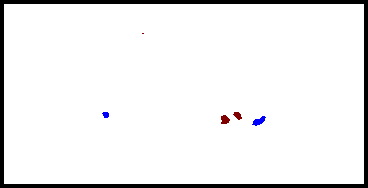}
    \label{subfig:gen_rem_good_missed}}
  \subfloat[New coronal holes that appear in model map.]
  { \includegraphics[width=0.47\textwidth]{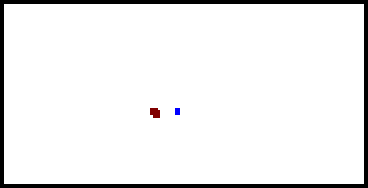}
    \label{subfig:gen_rem_good_new}}
  \caption{An example that demonstrates  good detection of missing and new coronal holes (07-13-2010). In this case, manual labeling and the algorithm agreed on all of the coronal holes 
    except for the upper-right coronal hole depicted in (d).}
  \label{fig:ch5genrem_good_case}
\end{figure}

\begin{figure}
  \subfloat[Consensus map.]
  { \includegraphics[width=0.47\textwidth]{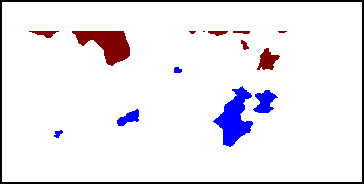}
    \label{subfig:gen_rem_bad_consensus}}
  \subfloat[Model map.]
  { \includegraphics[width=0.47\textwidth]{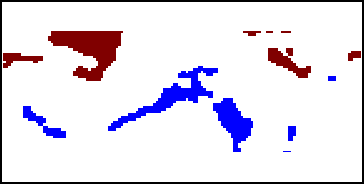}
    \label{subfig:gen_rem_bad_model}}

  \subfloat[Coronal holes missing from model.]
  { \includegraphics[width=0.47\textwidth]{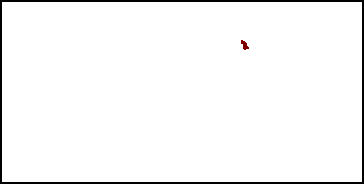}
    \label{subfig:gen_rem_missed}}
  \subfloat[New coronal holes in model.]
  { \includegraphics[width=0.47\textwidth]{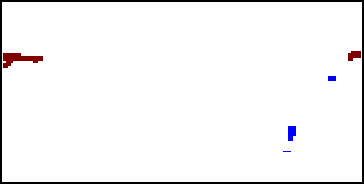}
    \label{subfig:gen_rem_new}}
  
  \subfloat[Manual matching.]
  { \includegraphics[width=0.94\textwidth]{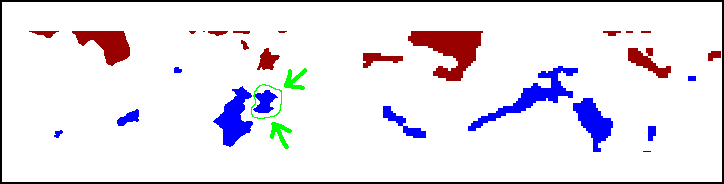}
    \label{subfig:gen_rem_man_matched}}
  \caption{A difficult coronal hole detection example (05-02-2011). The coronal hole shown in green is manually classified as removed but it is  classified as matched by the algorithm.}
  \label{fig:ch5genrem_bad_case}
\end{figure}

We present two detection examples.
An example where everything worked well is shown in Fig. \ref{fig:ch5genrem_good_case}.
On the other hand, a difficult case is shown in Fig. \ref{fig:ch5genrem_bad_case}. 
Visual inspection of the coronal holes of Fig. \ref{fig:ch5genrem_good_case}
demonstrates that the physical model map matched the consensus map
and did not produce a large number of new and missing coronal holes.
On the other hand, there was much more activity and there were many
differences in the maps of Fig. \ref{fig:ch5genrem_bad_case}.

\section{Matching}\label{sec:matching}
\begin{figure}
  \subfloat[Clustered, positive polarity consensus map.]
  { \includegraphics[width=0.47\textwidth]{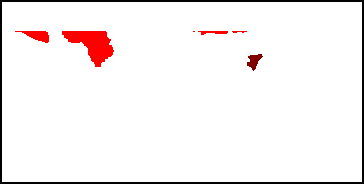}}
  ~~\subfloat[Clustered, positive polarity model map.]
  { \includegraphics[width=0.47\textwidth]{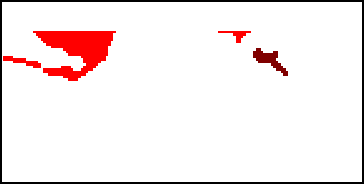}}
  
  \subfloat[Matching clusters from Consensus (left) and model (right) for positive polarity.]
  { \includegraphics[width=0.94\textwidth]{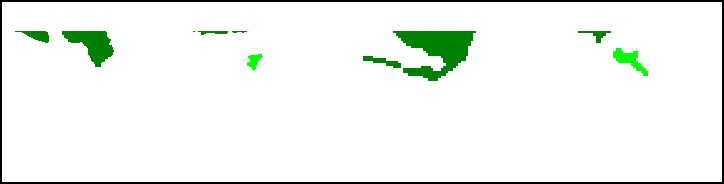}}

  \subfloat[Clustered, negative polarity consensus map.]
  { \includegraphics[width=0.47\textwidth]{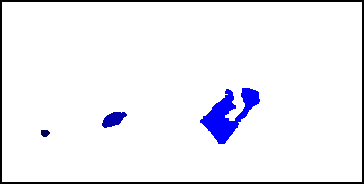}}
  ~~\subfloat[Clustered, negative polarity model map.]
  { \includegraphics[width=0.47\textwidth]{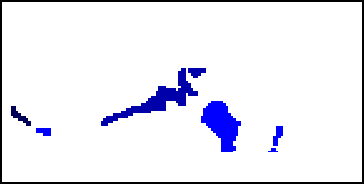}}        
  
  \subfloat[Matching clusters from consensus (left) and model (right) for negative polarity.]
  { \includegraphics[width=0.94\textwidth]{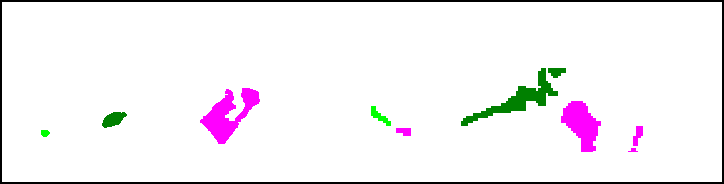}}
  
  \caption{Good matching example (02-04-2011). Matched clusters are shown using the same colors.}
  \label{fig:ch5Matching_good}
\end{figure}

\begin{figure}
  \subfloat[Clustered, positive polarity consensus map.]
  { \includegraphics[width=0.47\textwidth]{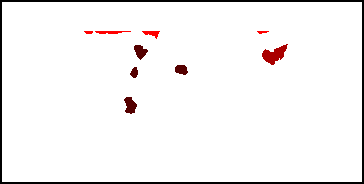}
    \label{subfig:ch5consensus_cluster}}
  ~~\subfloat[Clustered, positive polarity  model map.]
  { \includegraphics[width=0.47\textwidth]{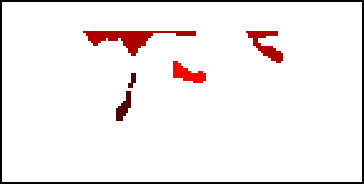}}        
  
  \subfloat[Matching clusters from consensus (left) and model (right) for positive polarity.]
  {\includegraphics[width=0.94\textwidth]{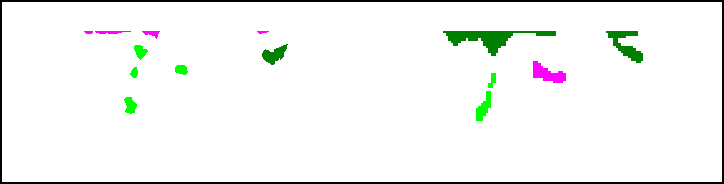}}
  
  \subfloat[Clustered, negative polarity consensus map.]
  { \includegraphics[width=0.47\textwidth]{pictures/new_chapter5/matching_example/20100807_1ref_pos_matchable_clus.png}}
  ~~\subfloat[Clustered, negative polarity model map.]
  { \includegraphics[width=0.47\textwidth]{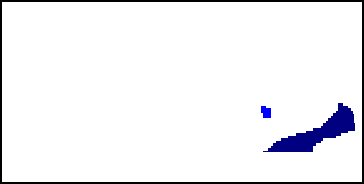}}        
  
  \subfloat[Matching clusters from consensus (left) and model (right) for negative polarity.]
  {\includegraphics[width=0.94\textwidth]{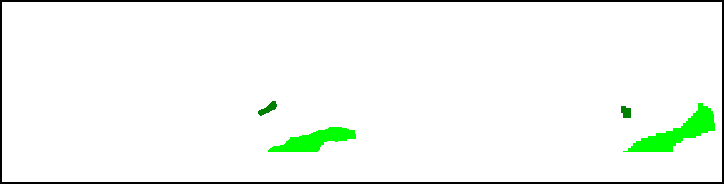}}
  
  \caption{Difficult example of coronal hole matching showing issues in the algorithm 
    (07-08-2010). Matched clusters are shown using same colors.}        
  \label{fig:ch5Matching_bad}
\end{figure}

Results of matching using linear programming are demonstrated in Figs. \ref{fig:ch5Matching_good} 
and \ref{fig:ch5Matching_bad}.
Fig. \ref{fig:ch5Matching_good} shows an example where automated matching
agrees with manual matching.
Fig. \ref{fig:ch5Matching_bad} shows an example where there are significant
differences between automated and manual matching (see pink cluster).

\section{Classification results}\label{sec:classification_results}
Map classification is performed using KNN and SVM classifiers.
Classifiers are trained on $10$ randomly chosen dates  and tested on remaining $40$ (as described in
Sec \ref{sec:ground_truth}). Table \ref{tab:ch5classificaiton_results_auto}
provides a side by side comparison of classification accuracy when
using consensus and automatically segmented maps as references.
We can see that the use of automatically segmented maps produced results
that approximated the use of consensus maps.

In table \ref{tab:ch5classificaiton_results_eachExpert}, we present results based on the use of the original, manual segmentations (\textit{\textbf{R4}} and \textit{\textbf{R7}}). 
We can see that \textit{\textbf{R7}} is closer to consensus maps than \textit{\textbf{R4}}.

\begin{table} 
  \begin{tabular}{l |l l  l l | l l  l l }
    & \multicolumn{4}{l|}{\bf Consensus Vs model} & \multicolumn{4}{p{4cm}}{\bf Auto segmentation Vs model}\\
    & \multicolumn{2}{c}{KNN} & \multicolumn{2}{c |}{SVM} & \multicolumn{2}{c}{KNN} & \multicolumn{2}{c}{SVM}\\
    \hline
    &  {\it good} & {\it bad} & {\it good} & {\it bad} & {\it good} & {\it bad} & {\it good} & {\it bad} \\
    {\it good}    & 143 & 46 & 29  & 36  & 121 & 12  & 111 & 15 \\
    {\it bad}     & 84 & 207 & 198 & 217 & 106 & 241 & 116  & 238\\
    \hline
    Accuracy & \multicolumn{2}{c}{$72.9\%$} & \multicolumn{2}{c |}{$51.2$} & \multicolumn{2}{c}{$75.4\%$} & \multicolumn{2}{c}{$72.7\%$}
  \end{tabular}
  \caption{Classification results 1}
  \label{tab:ch5classificaiton_results_auto}
\end{table}

\begin{table}
  \begin{tabular}{l |l l  l l   |    l l  l l }
    &\multicolumn{4}{l|}{\bf \textit{\textbf{R4}} Vs model}& \multicolumn{4}{p{4cm}}{\bf \textit{\textbf{R7}} Vs model}\\
    &       \multicolumn{2}{c}{KNN} & \multicolumn{2}{c |}{SVM} & \multicolumn{2}{c}{KNN} & \multicolumn{2}{c}{SVM}\\
    \hline
    &{\it good}&{\it bad}&{\it good}&{\it bad}&{\it good}&{\it bad}&{\it good}&{\it bad}\\
    {\it good} & 78  & 88 & 55 & 33 & 74 & 39 & 34 &15\\
    {\it bad}  & 149 & 165 & 172 & 220 & 141 & 214 & 181 & 238\\
    \hline
    Accuracy & \multicolumn{2}{c}{$50.6\%$} & \multicolumn{2}{c |}{$57.2\%$} & \multicolumn{2}{c}{$61.5\%$} & \multicolumn{2}{c}{$58.1\%$}
  \end{tabular}
  \caption{Classification results 2}
  \label{tab:ch5classificaiton_results_eachExpert}
\end{table}

\chapter{Conclusion}
\section{Thesis overview}
The primary contributions of the thesis include: 
(i) a new segmentation method to detect coronal holes, 
(ii) a manual protocol to support reproducible classification of physical models, and
(iii) an automated method for physical model classification.
In each case, the performance of each method was validated against 
human experts.
By comparing against the consensus maps,
the new coronal hole segmentation method was shown to work better
than the currently used method.
Similarly, when compared against consensus maps,
automated classification of the physical maps has been
shown to be equal or better than what can be achieved by individual raters.
Here, it is important to note that performance cannot exceed what can be done
by the consensus maps, since these maps are used as the ground truth
for the segmentation methods.  

\section{Future Work}
There is big room for improvement in the proposed methods.
A summary of the most important issues includes:
\begin{itemize}
\item \textbf{Larger number of manual segmentation by independent raters.}
  There were significant differences between the consensus maps and the individual
  segmentation maps produced by each rater. Such issues can only be addressed by including
  a substantially larger number of raters. In this case, if nothing else, the collection
  of all of the maps that will be produced will likely not missed any critical coronal holes.
\item \textbf{Inter-rater and intra-rater variability studies.} Based on a larger number of 
  raters, we can measure the statistical variation between them, as well as the variation
  when the same rater repeats the process.
\item \textbf{Studies on larger databases.} Clearly, it will be interesting to extend
  the study to cover more dates.
\item \textbf{Prediction studies.} Instead of looking at standard solar cycles, it will 
  be interesting to develop prediction methods that use previously processed maps
  to classify physical maps in future dates.
\item \textbf{Extracting coronal holes from the Helios events knowledgebase (HEK).}
  It will be interesting to repeat the study by using HEK maps to generate consensus maps
  and then repeat the study based on the new consensus maps.                               
\end{itemize}

\begin{comment}
  % 
  % Chapter-- How to write
  % 
  \chapter{How to write}
  \section{Dr. Pattichis}
  \begin{itemize}
  \item When writing the format should be something understandable to average reader.
  \item Never refer forward.
  \item What is not obvious and innovations are the two things that needs focus.
  \item You can have what ever you want in appendix as long as it is sufficiently covered
    in main body.
  \item We are image processing group, so use figures to tell the story.
  \end{itemize}
  \section{Latex personal documentation}
  \begin{itemize}
  \item Drawing subfigures
  \item Calling a function in algorithm
  \item Dividing figure between two pages.
  \item rotating figure 90 degrees.
  \item complicated table example.
  \item in a figure caption comes before label.
  \end{itemize}
\end{comment}

% Bibliography
\nocite{*}
\bibliographystyle{plain}       %???, instead of 'plain' should use 'AMS'. but
\bibliography{thBib}           % it did not work so moving on with plain

\end{document}